\documentstyle[12pt]{article}
\begin{document}

\date{}

\title{Gravity, torsion, Dirac field and computer algebra using MAPLE 
and REDUCE}

\author{Dumitru N. Vulcanov\thanks{
Permanent address : The West University of Timi\c soara,
Theoretical and Computational Physics Department,
B-dul V. P\^ arvan no. 4, 1900 Timi\c soara,  Rom\^ ania, e-mail : 
{\tt vulcan@physics.uvt.ro}}\\
Max-Planck-Institut f\" ur Gravitationsphysik\\
Albert-Einstein-Institut\\
Numerical Relativity Group\\
Golm, Am M\" uhlenberg 1, D-14476, Germany}

\maketitle

\begin{abstract}

The article presents computer algebra procedures and routines  applied
to the study of the Dirac field on curved spacetimes. The main part of 
the procedures is 
devoted to the construction of Pauli and Dirac matrices algebra on an 
anholonomic orthonormal reference frame. Then these procedures are used
to compute the Dirac equation on curved spacetimes in a sequence of
special dedicated routines. A comparative review of such procedures 
obtained for two computer algebra platforms (REDUCE + EXCALC and MAPLE + 
GRTensorII) is carried out. Applications for the calculus of Dirac equation 
on specific examples of spacetimes with or without torsion are pointed out.

\end{abstract}



\section{Introduction} 

\noindent In a series of recent articles \cite{12}-\cite{13}
 we have presented some routines and their applications, written in 
REDUCE+EXCALC computer algebra language for performing calculations 
in Dirac theory on curved spacetimes. Including the Dirac fields in 
gravitation theory requires lengthy (or cumbersome) calculations which 
could be solved by computer algebra methods. Initially our main purpose 
was to develop a complete algebraic programming package for this purpose 
using only the REDUCE + EXCALC platform. This program was accomplished 
and the main results and applications were reported in our above cited 
articles \cite{12}-\cite{13}. But we are aware of the fact that other very 
popular algebraic manipulations systems are on the market (like Mathematica 
or MAPLE) thus the area of people interested in algebraic programming 
routines for Dirac equation should be much larger. In this perspective 
we developed similar programs and routines for MAPLE \cite{14} platform
using the package GRTensorII \cite{15} (adapted for doing calculations 
in General Relativity). Because there is no portability between these two
systems we were forced to compose completely new routines, in fact following
the same strategy I used in REDUCE : first, the Pauli and Dirac matrices 
algebra (using only the MAPLE environment) and then the construction of 
the Dirac equation on curved spacetimes where the capabilities of GRTensorII 
package is used. Because the authors of GRTensorII offer also package 
versions  for Mathematica we can be sure that our  MAPLE routines could be 
easily adapted for Mathematica which would  highly increase the number of 
users of our procedures.

This article is organized as follows : the next section no. 2 presents a 
short review of the theory of Dirac fields on curved spacetime,  pointing 
out the main notations and conventions we shall use through the article. 
Then  section no. 3 is devoted to a short overview of our routines and 
programs in REDUCE+EXCALC previously described in great detail in \cite{12}. 
This section is necessary in view of the fact that I applied the same 
strategy for constructing our programs in MAPLE as is pointed out above. 
Section no. 4 contains a complete description of our programs in 
MAPLE+GRTensorII. We also included here some facts about the main 
differences (advantages and disadvantages) between the two algebraic 
programming platforms (REDUCE and MAPLE). Section no. 5 is devoted to 
the problem of including spacetimes with torsion in order to compute the
Dirac equation using our MAPLE procedures. The last section of the article
includes a list of some of spacetimes examples we used in order to 
calculate the Dirac equation. Two of these examples are spacetimes with
torsion thus it is  pointed out the contribution of torsion components to the
Dirac field. Several applications of our programs (in REDUCE or in MAPLE) are
to be developed in our future projects : searching for inertial effects in 
non-inertial systems of reference (partially presented in \cite{13} for a 
Schwarzschild metric without torsion) or quantum effects (as in \cite{10}) 
in order to provide new theoretical results  for  experimental gravity 
\cite{16}.

Through the article we  use standard notations \cite{1},\cite{2},
\cite{3}, \cite{17}. 
Four dimensional spacetime indices will run from 0 to 3 and are denoted
with Greek letters  while spatial three-dimensional indices are denoted
with the Latin  ones. The computer algebra system used  was REDUCE 3.6 
\cite{4} with EXCALC package \cite{5} or MAPLE V \cite{14} with GRTensorII 
package \cite{15}.

\section{Pauli and Dirac matrices algebra and Dirac equation on curved 
space-time}

The main problem is to solve algebraic expressions involving the 
Dirac matrices \cite{1},\cite{2}, \cite{17}. 
To this end it is convenient to construct explicitly  
these matrices as a direct product of several pairs among the Pauli matrices
$\sigma_i, i = 1,2,3$, and the $2 \times 2$ unit matrix. Thus all the 
calculations will be expressed in terms of the Pauli matrices and 
2-dimensional 
Pauli spinors. Consequently the  result will be obtained in a form 
which is suitable for physical interpretations. We shall consider the Pauli 
matrices as abstract objects with specific multiplication rules. 
Thus we work
with operators instead of their matrices in a spinor representation. However,
if one desires to see the result in the standard Dirac form with $\gamma$
matrices 
it will be sufficient to use a simple reconstruction procedure which will be
presented in the next section \cite{12}. 

We shall consider the Dirac formalism 
in the chiral form where the Dirac matrices are \cite{2} :
\begin{equation}\label{1}
\gamma^0 = \left [ \begin{array}{cc} 0 & 1 \\ 1 & 0 \end{array} \right ]
\hbox{~~,~~}
\gamma^i = \left [ \begin{array}{cc} 0 & \sigma_i \\ -\sigma_i & 0 \end{array}
\right ]\hbox{~~,~~}
\gamma^5 = \left [ \begin{array}{cc} -1 & 0 \\ 0 & 1 \end{array}
\right ]
\end{equation}
The Dirac spinor :
\begin{equation}\label{2}
\Psi = \left [ \begin{array}{cc} \varphi_l \\ \varphi_r \end{array} \right ] 
\in {\cal{H}}_D
\end{equation}
involves the Pauli spinors $\varphi_l$ and $\varphi_r$ which transform
according to the irreducible representations $(1/2,0)$ and $(0,1/2)$ of the
group SL(2,C). In this representation the left and right-handed Dirac 
spinors are
\begin{equation}\label{3}
\Psi_L = \frac{1-\gamma^5}{2} \Psi = \left [ 
          \begin{array}{c} \varphi_l \\ 0 \end{array} \right ] \hbox{~~,~~}
\Psi_R = \frac{1+\gamma^5}{2} \Psi = \left [ 
          \begin{array}{c} 0 \\ \varphi_r \end{array} \right ] 
\end{equation}
and, therefore, the Pauli spinors $\varphi_l$ and $\varphi_r$ will be the
left and the right-handed parts of the Dirac spinor. The SL(2,C) generators
are 
\begin{equation}\label{4}
S^{\mu\nu} = \frac{i}{4} \left [ \gamma^{\mu},\gamma^{\nu}\right ]
\end{equation}
It is shown \cite{1} that ${\cal{H}}_D = {\cal{H}} \otimes {\cal{H}}$ ( where 
${\cal{H}}$ is the two-dimensional space of Pauli spinors) and, therefore the
Dirac spinor can be written as:
\begin{equation}\label{5}
\Psi = \xi_1 \otimes \varphi_l + \xi_2 \otimes \varphi_r 
\hbox{~~with~~} \xi_1 = \left [ \begin{array}{c} 1 \\ 0 \end{array} \right ]
\hbox{~~and~~} \xi_2 = \left [ \begin{array}{c} 0 \\ 1 \end{array} \right ]
\end{equation}
while the $\gamma$-matrices and the SL(2,C) generators can be put in the form :
\begin{equation}\label{6}
\gamma^0 = \sigma^1 \otimes 1 \hbox{~~,~~} 
\gamma^k = i \sigma^2 \otimes \sigma^k \hbox{~~,~~}
\gamma^5 = - \sigma^3 \otimes 1 
\end{equation}
\begin{equation}\label{7}
S^{ij} = \frac{1}{2}\epsilon_{ijk} 1 \otimes \sigma^k \hbox{~~,~~} 
S^{0k} = - \frac{i}{2}\sigma^3 \otimes \sigma^k 
\end{equation}
These properties allow to reduce the Dirac algebra to that of the Pauli
matrices.

In order to introduce Dirac equation on curved space-time we have always used 
an anholonomic orthonormal frame because at any point of spacetime we need 
an {\it orthonormal reference frame} in order to describe the spinor field 
as it is already pointed before \cite{12}).
The Dirac equation in a general reference of frame, defined by an anholonomic 
tetrad field is \cite{7} :
\begin{equation}\label{8}
i\hbar\gamma^{\mu}D_{\mu}\Psi = mc \Psi 
\end{equation}
where the covariant Dirac derivative $D_{\mu}$ is 
\begin{equation}\label{9}
D_{\nu}=\partial_{\nu} +
  \frac{i}{2}S^{\rho \mu}\Gamma_{\nu \rho \mu} 
\end{equation}
and where $S^{\mu \nu}$ are the SL(2,C) generators (\ref{4}) and $\Gamma_{\nu \rho \mu}$
are the anholonomic components of the connection.

\section{Review of the REDUCE+EXCALC routines for calculating the Dirac equation}

We shall describe here those part of the program realizing the Pauli and Dirac 
matrix algebras \cite{12},\cite{13}. In the first lines of this sequence 
we introduce the operators and the non-commuting operators being 
useful throughout
the entire program. The Pauli matrices are represented using the operator 
{\bf p} with one argument. The Dirac matrices are denoted by {\bf gam} 
of one argument (an operator if we use only REDUCE, or for EXCALC package it 
will be a 0-form with one index) while the operator {\bf dirac} stands 
for the Dirac equation. The SL(2,C) generators (\ref{4}) are denoted by the 
0-form {\bf s(a,b)}.The 
basic algebraic operation, the commutator ({\bf com}) and anticommutator 
({\bf acom}) are then defined here only for commuting (or anticommuting) 
only simple objects (``kernels''). For commuting more complex expressions, 
(in order to introduce some necessary commutation relations) a more complex 
operator is necessary to  introduce. 
Other objects, having a more or less local utilization in the program will 
be introduced with  declarations and statements at their specific appearance.

The main part of the program is the Pauli subroutine : 
\begin{verbatim}
        LET p(0)=1;                           
        LET p(2)*p(1)=-p(1)*p(2);             
        LET p(1)*p(2)=i*p(3);                 
        LET p(3)*p(1)=-p(1)*p(3);             
        LET p(1)*p(3)=-i*p(2);               
        LET p(3)*p(2)=-p(2)*p(3);            
        LET p(2)*p(3)=i*p(1);                 
        LET p(1)**2=1;                        
        LET p(2)**2=1;                        
        LET p(3)**2=1;                       
\end{verbatim}
The Pauli matrices, $\sigma_i$ appear as {\bf p(i)} while the $2 \times 2$
unity matrix is {\bf p(0)=1}. The properties of the Pauli matrices are
given by the above sequence of 10 lines.
The {\bf direct product} denoted by the {\bf pd} operator has properties, 
introduced as :
\begin{verbatim}
for all a,b,c,u let pd(a,b)*pd(c,u)=pd(a*c,b*u);   
for all a,b let pd(a,b)**2=pd(a**2,b**2);          
for all a,b let pd(-a,b)=-pd(a,b);                 
for all a,b let pd(a,-b)=-pd(a,b);                 
for all a,b let pd(i*a,b)=i*pd(a,b);               
for all a,b let pd(a,i*b)=i*pd(a,b);               
for all a let pd(0,a)=0;                           
for all a let pd(a,0)=0;                           
let pd(1,1)=1;                                     
\end{verbatim}
\noindent  Some difficulties arise from the bilinearity of the direct 
product which
requires to identify all the scalars involved in the current calculations. 
This can be done only by using complicated procedures or special assignments.
For this reason we shall use a special definition of the direct product
({\bf pd}) which gives up the general bilinearity property. The operator 
{\bf pd} will depend on two Pauli matrices or on the Pauli matrices with 
factors $-1$ or $\pm i$. It is able to recognize only these numbers but this
is enough since the multiplication of the Pauli matrices has just the 
structure constants $\pm 1$ and $\pm i$ (we have $\sigma_i\sigma_j = 
\delta_{ij}+ i\epsilon_{ijk}\sigma_k$).

Thus by introducing the multiplication rules of the direct product it will 
be sufficient to give some instructions (see above) which represent the 
bilinearity, defined only for the scalars $-1$ and $i$. The next two 
instructions represent the definition of ``$0$'' in the direct product space, 
while the last one from the above sequence introduces the $4 \times 4$ unit 
matrix. 

Thus the $\gamma$-matrices can be defined now with the help of our direct 
product; we added also here the definition of the SL(2,C) generators from 
eq. (\ref{4}) :
\begin{verbatim}
gam(1):=i*pd(p(2),p(1));           
gam(2):=i*pd(p(2),p(2));            
gam(3):=i*pd(p(2),p(3));      %  Remember that gam's are
gam(0):=  pd(p(1),1);         %        0-forms !!!
gam5  := -pd(p(3),1);         %  instead of ``gam(5)''

s(a,b):= i*com(gam(a),gam(b))/4;                    
\end{verbatim}

 All the above program lines we have presented here can be 
used for the Dirac theory on the Minkowski space-time in an inertial system of 
reference. Here we shall point at first the main differences which appear 
when one wants to run our procedures on some curved space-times or in a 
non-inertial reference of frame. Some of these minor differences are already 
integrated in the lines presented in the previous section.

First of all we have to add, at the beginning of the program some EXCALC lines
containing the metric statement.  We have always used an anholonomic 
orthonormal frame because at any point of spacetime we need an {\it 
orthonormal reference frame} in order to describe the spinor field as it has
been already pointed before \cite{12}).

After the above metric statement we must add in the program
all the procedures described in the last section. Then we can
introduce some lines calculating the Dirac equation in this context.
As a result we have used the next sequence of EXCALC lines :
\begin{verbatim}
pform {der(j),psi}=0; fdomain psi=psi(x,y,z,t);           
der(-j):= ee(-j)_|d psi + (i/2)*s(b,h)*cris(-j,-b,-h);    

operator derp0,derp1,derp2,derp3;                         
noncom derp0,derp1,derp2,derp3;                           

let @(psi,t)=derp0;                                       
let @(psi,x)=derp1;                    
let @(psi,y)=derp2;
let @(psi,z)=derp3;                                       

dirac := i*has*gam(j)*der(-j)-m*c;                        
ham:= -(gam(0)*(1/(ee(-0)_|d t))*dirac-i*has*derp0);      
\end{verbatim}
In defining the Dirac derivative {\bf der} we have introduced also an 
formal Dirac spinor ({\bf psi}) being a 0-form and depending on the 
variables imposed by the symmetry of the problem. It is just an intermediate
step (in fact a ``trick'') in order to obtain the partial derivative 
components as operators, because after calculating the components of the
covariant derivative ({\bf der(-j)}- see above) we have to replace
the partial derivatives of {\bf psi} with four non-commuting operators
{\bf derp0}, {\bf derp1} ... {\bf derp3}. 
The Dirac operator is thus defined as {\bf dirac}$:= (i \hbar \gamma^{\mu}
D_{\mu} - mc)\Psi$ and finally the Dirac Hamiltonian ({\bf ham}) is 
obtained from the canonical form of the Dirac equation :
$$i\hbar \frac{\partial \psi}{\partial t} = H \psi $$
which we shall use later, in the study of the nonrelativistic approximation
of the Dirac equation in non-inertial reference frames. 

The results we have obtained after processing the program lines 
presented until now contain only the Pauli matrices and direct products of 
Pauli matrices. If one wants to have the final result in terms of the 
$\gamma$-matrices and SL(2,C) generators (and not in terms of direct products
of Pauli matrices) the procedure {\bf rec} should be used :
\begin{verbatim}
operator gama,gen;
noncom gama,gen;
PROCEDURE rec(a);
begin;
          ws1:=sub(pd(p(1),1)=gama(0),a);
          ws1:=sub(pd(p(2),1)=-i*gama(0)*gama(5),ws1);
          ws1:=sub(pd(p(3),1)=-gama(5),ws1);
          ws1:=sub(pd(1,p(1))=2*gen(2,3),ws1);
          ws1:=sub(pd(1,p(2))=2*gen(3,1),ws1);
          ws1:=sub(pd(1,p(3))=2*gen(1,2),ws1);
     for k:=1:3 do 
        <<ws1:=sub(pd(p(2),p(k))=-i*gama(k),ws1);
          ws1:=sub(pd(p(1),p(k))=gama(k)*gama(5),ws1);
          ws1:=sub(pd(p(3),p(k))=2*i*gen(0,k),ws1)>>;
   return ws1;
end;
\end{verbatim}
\noindent This is an operator depending on an expression involving matrices
({\bf a}) which reconstructs the $\gamma$-matrices and the SL(2,C) generators 
from the direct products of Pauli matrices according to eqs. (6) and (7). 

We must point out  that the new introduced operators {\bf gama} and {\bf gen} 
do not represent a complete algebra. They are introduced in order to have a 
result in a comprehensible form. Thus, in  this form  the result cannot be 
used in further computations. Only the results obtained
before processing the {\bf rec} procedure could be used, in order to benefit
of the complete Pauli and Dirac matrices algebra.

\section{MAPLE+GRTensorII procedures for calculating the Dirac equation}

Here we shall present, in details, our procedures in MAPLE+GRTensorII for
calculating the Dirac equations, pointing out the main differences between
MAPLE and REDUCE programming in obtaining the same results. The first major
problem appears in MAPLE when one try to introduce the Pauli and Dirac matrices
algebra. In MAPLE this will be a difficult task because the ordinary product
(assigned in MAPLE with ``$*$'') of operators is automatically commutative,
associative, linear, etc. like an ordinary scalar product - in REDUCE these
properties are active only if the operators are declared previously as having
such properties. Thus we have to define two special product operators :
for Pauli matrices $\sigma_{\alpha}, \alpha=0,1,2,3$ 
(assigned in our procedures with {\bf pr}) 
and for the direct product of Pauli matrices (assigned here also with 
the operator {\bf pd(``,``)}) which is assigned with ``{\bf \&p}''. As a 
consequence we have to introduce long lists with their properties as, 
for example :
\begin{verbatim}
> define(sigma,sigma(0)=1);

> define(pr,pr(1,1)=1,pr(1,sigma(1))=sigma(1),pr(1,sigma(2))=
sigma(2),pr(1,sigma(3))=sigma(3),pr(sigma(1),1)=sigma(1),
pr(sigma(2),1)=sigma(2),pr(sigma(3),1)=sigma(3),...
pr(sigma(1),sigma(1))=1,pr(sigma(2),sigma(2))=1,...

> define(pd,pd(0,a::algebraic)=0,pd(a::algebraic,0)=0,pd(1,1)=1,
pd(I*a::algebraic,b::algebraic)=I*pd(a,b),
pd(-I*a::algebraic,b::algebraic)=-I*pd(a,b),
pd(a::algebraic,I*b::algebraic)=I*pd(a,b),
pd(a::algebraic,-I*b::algebraic)=-I*pd(a,b),
pd(-a::algebraic,b::algebraic)=-pd(a,b),
pd(a::algebraic,-b::algebraic)=-pd(a,b));

> define(`&p`,`&p`(-a::algebraic,-b::algebraic)=`&p`(a,b),...
\end{verbatim}
Of course the reader may ask why is 
not much simpler to declare, as an example, the {\bf \&p} as being linear (or 
multilinear) ? Because in this case the operator does not act properly, the 
linearity property  picking out from the operator all the terms, being or not
Pauli matrices or direct products {\bf pd} of Pauli matrices. Thus it is 
necessary to forget the linearity and to introduce, as separate properties 
all the possible situations to appear in the calculus. As a result the
program becomes very large with a corresponding waste of RAM memory and speed
of running. This will be the main disadvantage of MAPLE version of our 
program in comparison with the short (and, why not, elegant) REDUCE procedures.
Of course, in a more compact version of our programs, we defined MAPLE routines
with these operators, and the user needs only to load at the beginning of 
MAPLE  session these routines, but there is no significant economy of 
memory and running time. 

The next step is to define Dirac $\gamma$-matrices and a special commutator
(with {\bf \&p}) :
\begin{verbatim}
> define(gam,
gam(1)=I*pd(sigma(2),sigma(1)),
gam(2)=I*pd(sigma(2),sigma(2)),
gam(3)=I*pd(sigma(2),sigma(3)),
gam(0)=pd(sigma(1),1),gam(5)=-pd(sigma(3),1));

> define(comu,comu(a::algebraic,b::algebraic)=a &p b - b &p a);
\end{verbatim}

The next program-lines are in GRTensorII environment, thus is necessary to
load first the GRTensorII package and then  the corresponding
metric (with {\bf qload(...)} command \cite{15}). It follows then :
\begin{verbatim}
> grdef(`SS{ ^a ^b }`);
> grcalc(SS(up,up)):
> (I/4)*comu(gam(0),gam(0));
> (I/4)*comu(gam(1),gam(0));
      .
      .
      .
> (I/4)*comu(gam(3),gam(3));
> grdisplay(SS(up,up));

> grcalc(Chr(dn,dn,dn));grdisplay(Chr(dn,dn,dn));
> grcalc(Chr(bdn,bdn,bdn));grdisplay(Chr(bdn,bdn,bdn));
\end{verbatim}
These are a sequence of commands in GRTensorII for defining the SL(2,C) 
generators $S^{ij}$ (as the tensor {\bf SS\{ \^{ }a \^{ }b\}}) using formula 
(\ref{4})
and for the calculus of Christoffel symbols in an orthonormal frame base 
({\bf Chr(bdn,bdn)}). Here it becomes obvious one of the main advantages of 
MAPLE+GRTensorII platform, namely, the possibility of computing of 
the tensor components both in a general reference frame or in an anholonomic 
orthonormal frame which is vital for our purpose of construction the  Dirac 
equation. 

Next we have to define, as two vectors  the Dirac-$\gamma$ matrices (assigned 
as the contravariant vector {\bf ga\{ \^{ }a \}} and the derivatives 
of the wave
function $\psi$ (assigned as the covariant vector {\bf Psid\{ a \}} in order
to use the facilities of GRTensorII for manipulating with indices :
\begin{verbatim}
> grdef(`ga{ ^a }:=[gam(0),gam(1),gam(2),gam(3)]`);
> grdisplay(ga(up));

> grdef(`Psid{ a }:=[diff(psi(x,t),t),diff(psi(x,t),x),
                      diff(psi(x,t),y),diff(psi(x,t),z)]`);
> grcalc(Psid(dn));grdisplay(Psid(dn));
> grcalc(Psid(bdn));grdisplay(Psid(bdn));
\end{verbatim}

The next step is to define the term $\frac{i}{2}
S^{\rho\mu}\Gamma_{\nu\rho\mu}$ from equation (\ref{9}) :
\begin{verbatim}
> grdef(`de{ i }:=(I/2)*SS{ ^a ^b }*Chr{ (i) (a) (b) }`); 
> grcalc(de(dn));grdisplay(de(dn));
\end{verbatim}
The components of {\bf de\{ i \}} are polynomials containing direct
products {\bf pd(...)} of Pauli matrices and the product
between $\gamma^{\nu}$ and  $\frac{i}{2}S^{\rho\mu}\Gamma_{\nu\rho\mu}$ 
from equation (\ref{8}) is, in fact, the special product {\bf \&p}.
Thus we  obtain the term $\gamma^{\nu} S^{\rho\mu}\Gamma_{\nu\rho\mu}$ 
(denoted below with the operator {\bf dd}) 
by a special MAPLE sequence which in fact split the components of 
{\bf de\{ i \}} in monomial terms and then execute the corresponding 
{\bf \&p} product, finally reconstructing the {\bf dd} operator :
\begin{verbatim}
 a0:=expand(grcomponent(de(dn),[t]));a00:=0;
> u0:=whattype(a0);u0; nops(a0);
> if u0=`+` then  for i from 1 to nops(a0) do 
     a00:=a00+I*h*grcomponent(ga(up),[t]) &p op(i,a0) od else 
     a00:=I*h*grcomponent(ga(up),[t]) &p a0 fi; a00;

> a1:=expand(grcomponent(de(dn),[x]));a11:=0;
> u1:=whattype(a1);u1;
> nops(a1);
> if u1=`+` then  for i from 1 to nops(a1) do 
     a11:=a11+I*h*grcomponent(ga(up),[x]) &p op(i,a1) od else 
     a11:=I*h*grcomponent(ga(up),[x]) &p a1 fi;a11;
      .
      .
      .
> grdef(`dd`); grcalc(dd);
> a00+a11+a22+a33;
> grdisplay(dd);
\end{verbatim}
Finally the Dirac equation is obtained as :
\begin{verbatim}
> grdef(`dirac:=I*h*ga{ ^l }*Psid{ (l) } + dd*psi(x,t) - m*c*psi(x,t)`);
> grcalc(dirac);
> grdisplay(dirac);
\end{verbatim}
In order to obtain the Dirac equation in a more comprehensible form we have the
next sequence of MAPLE commands (similar to the reconstruction {\bf rec} 
procedure from the REDUCE program) :
\begin{verbatim}
> define(`gen`);
> define(`gama`);
> grmap(dirac,subs,pd(sigma(1),1)=gama(0),pd(sigma(2),1)=-I*gama(0)*
gama(5),pd(sigma(3),1)=-gama(5),pd(1,sigma(1))=2*gen(2,3),pd(1,
sigma(2))=2*gen(3,1),pd(1,sigma(3))=2*gen(1,2),pd(sigma(2),sigma(1))=
-I*gama(1),pd(sigma(2),sigma(2))=-I*gama(2),pd(sigma(2),sigma(3))=
-I*gama(3),pd(sigma(1),sigma(1))=gama(1)*gama(5),pd(sigma(1),sigma(2))=
gama(2)*gama(5),pd(sigma(1),sigma(3))=gama(3)*gama(5),pd(sigma(3),
sigma(1))=2*I*gen(0,1),pd(sigma(3),sigma(2))=2*I*gen(0,2),pd(sigma(3),
sigma(3))=2*I*gen(0,3),`x`);
> grdisplay(dirac);
\end{verbatim}
where, of course, as in the REDUCE version, the operators {\bf gen} and 
{\bf gama} do not represent a complete algebra.

\section{Dirac equation on spacetimes with torsion}

We shall present here the way we adapted  our MAPLE+GRTensorII programs 
in order to calculate the Dirac equation on space-times with torsion. 

The geometrical frame for General Relativity is a Riemannian  
space--time but one very promising generalization is the Riemann--Cartan 
geometry  which (i) is the most natural generalization of a Riemannian 
geometry by allowing a non--symmetric metric--compatible connection, (ii) 
treats spin on the same level as mass as it is indicated by the group  
theoretical analysis of the Poincar\'e group, and (iii) arises in most  
gauge theoretical approaches to General Relativity, as e.g. in the  
Poincar\'e--gauge theory  or supergravity \cite{6},\cite{7}.
However, till now there is no experimental evidence for torsion. 
On the other hand, from the lack of effects which may be due to  
torsion one can calculate estimates on the maximal strength of the  
torsion fields \cite{10}. 
In this aspect we think that it is possible, using computer
algebra facilities to approach new theoretical aspects on matter fields 
(for example the Dirac field) behavior on spacetimes with torsion in order
to point out new gravitational effects and experiments at microscopic level.

A metric compatible connection  in a Riemann-Cartan theory is 
related to the torsion components by (see \cite{6} - eq. (1.18))
\begin{equation}\label{tor1}
\Gamma_{\alpha \beta \gamma} = \tilde{\Gamma}_{\alpha \beta \gamma} -
\frac{1}{2}\left[ (C_{\alpha \beta \gamma} - C_{\beta \gamma \alpha} +
C_{\gamma \alpha \beta}) -
(T_{\alpha \beta \gamma} -T_{\beta \gamma \alpha} + T_{\gamma \alpha \beta})
\right]
\end{equation}
where $\tilde{\Gamma}_{\alpha \beta \gamma}$ are the components of the 
riemannian connection, $C_{\alpha \beta\gamma}$ is the object of anholonomicity
and $T_{\alpha \beta \gamma}$ are the components of the torsion.
The idea is to replace the connection components from
the covariant derivative appearing in eqs. (\ref{8}-\ref{9}) 
with the above ones,
of course after calculating them in an orthonormal anholonomic reference frame,
suitable for calculation of the Dirac equation. Thus the sequence for 
calculating the {\bf de\{dn\}} operator (see above) should be replaced with
\begin{verbatim}
> grdef(`de{ i }:=(I/2)*SS{ ^a ^b }*(CHR{ (i) (a) (b) }+
                   (1/2)*(tor{ (i) (a) (b) } -tor{ (a) (b) (i) } 
                                    + tor{ (b) (i) (a) })))`);
> grcalc(de(dn)); grdisplay(de(dn));
\end{verbatim}
where the new connection components $\bf CHR\{a,b,c\}$ are now defined by the
sequence
\begin{verbatim}
> grdef(`ee{ a ^b }:= w1{ ^b }*kdelta{ a $x } 
                       + w2{ ^b }*kdelta{ a $y } + 
          w3{ ^b }*kdelta{ a $z } + w4{ ^b }*kdelta{ a $t }`); 
> grcalc(ee(dn,up));
> grdisplay(ee(dn,up));
> grdef(`CC{ a b c }:=2*ee{ a ^i }*ee{ b ^ j }*ee{ c [j ,i] }`);
> grcalc(CC(dn,dn,dn)); grdisplay(CC(dn,dn,dn));
> grdef(`CHR{ (a) (b) (c) } := Chr{ (a) (b) (c) } - 
                                    (1/2)*(CC{ a b c } - 
                                CC{ b c a } + CC{c a b })`);
> grcalc(CHR(bdn,bdn,bdn)); grdisplay(CHR(bdn,bdn,bdn));
\end{verbatim}
and the rest of the routines are unchanged. The only problem remains now to
introduce in an adequate way the components of the torsion tensor.I used
the suggestion from \cite{6}, 
pointing that we can assume that the 2-form $T^{\alpha}$
associated to the torsion tensor should have the same pattern as the $d\theta^{
\alpha}$'s where $\theta^{\alpha}$ is the orthonormal coframe, who's
components are denoted in GRTensorII with {\bf w1\{dn\} ...w4\{dn\}}. Thus
this operation is  possible only after we introduced the metric (with
{\bf qload} command). Then calculating the derivatives of the orthonormal frame
vector basis components we can introduce the torsion components by 
inspecting carefully these derivatives. Here there is an example of how this 
becomes possible in MAPLE+GRTensorII using one of the metric examples 
presented in the next section:
\begin{verbatim}
> grcalc(w1(dn,pdn));grcalc(w2(dn,pdn));
> grcalc(w3(dn,pdn));grcalc(w4(dn,pdn));
> grdisplay(w1(dn,pdn)); grdisplay(w2(dn,pdn));
> grdisplay(w3(dn,pdn)); grdisplay(w4(dn,pdn));
> grcalc(w1(bdn,pbdn));grcalc(w2(bdn,pbdn));
> grcalc(w3(bdn,pbdn));grcalc(w4(bdn,pbdn));
> grdisplay(w1(bdn,pbdn));grdisplay(w2(bdn,pbdn));
> grdisplay(w3(bdn,pbdn));grdisplay(w4(bdn,pbdn));
> grdef(`tor{ ^a b c }:=w1{ b ,c }*kdelta{ ^a $x } + 
                        w2{ b ,c }*kdelta{ ^a $y } + 
                        w3{ b ,c }*kdelta{ ^a $z } + 
                        w4{ b ,c }*kdelta{ ^a $t }`);
> grcalc(tor(up,dn,dn));grdisplay(tor(up,dn,dn));
> grmap(tor(up,dn,dn),subs,f(x)=v4(x),h(x)=v3(x),g(x)=v2(x),`x`);
> grdisplay(tor(up,dn,dn));
> grcalc(tor(bup,bdn,bdn));
> grdisplay(tor(bup,bdn,bdn));
\end{verbatim}
The reader can observe that we first assigned the components of the torsion 
tensor (here denoted with {\bf tor\{up,dn,dn\}}) then after displaying his
components we can decide to substitute new functions describing the torsion
instead of the functions describing the metric. Of course finally we calculate
the components of the torsion in an orthonormal anholonomic reference frame
({\bf tor\{bup,bdn,bd\}}).

\section{Some specific results}

This section is devoted to a list of some recent results we obtained by
running our procedures in MAPLE+GRTensorII, already described in the previous
sections. First we tested our programs by re-obtaining the form of the Dirac
equations in several spacetime metrics, obtained with REDUCE+EXCALC
procedures and reported in our previous articles \cite{12}-\cite{13}. 
The concordance of these
results with previous ones was a good sign for us to proceed with more
complicated and new examples, including ones with torsion. Here we shall
present some of these late examples. \\

\noindent {\bf 1.} {\it Conformally static metric with  $\Phi$ and $\Sigma$
constant} \cite{6} where the line element is :
\begin{eqnarray}
ds^2 = e^{2\Phi t+2\Sigma}a(r)^2 dr^2 +e^{2\Phi t+2 \Sigma}r^2 d\theta^2 
+e^{2\Phi t+2 \Sigma}r^2~sin(\theta)^2 d\phi^2 -
e^{2\Phi t+2\Sigma}b(r)^2 dt^2 \nonumber
\end{eqnarray}
Thus the Dirac equation becomes :
\begin{eqnarray}
i\hbar e^{-\Phi t - \Sigma}\left[ \gamma^1 \left(\frac{1}{a(r)}\frac{\partial}{
\partial t} -\frac{1}{2a(r)b(r)}b^{\prime}(r) -\frac{1}{a(r)r}\right) 
+\gamma^2
\left(\frac{1}{r}
\frac{\partial}{\partial r} + \frac{1}{r~sin(\theta)}\frac{\partial}{\partial 
\theta} \right . \right .\nonumber\\
\left . \left .
- \frac{1}{2r}~cotg(\theta)\right) 
+ \frac{1}{b(r)} \gamma^0 \left( \frac{3}{2}
- \frac{\partial}{\partial \phi}\right)\right]\psi - mc\psi=0\nonumber
\end{eqnarray}
where $b^{\prime}(r)$ is the derivative $\partial b(r)/\partial r$.\\
 
\noindent {\bf 2.} {\it Taub-NUT spacetime} having the line element as :
\begin{eqnarray}
ds^2 = -4 l^2 U(t)dy^2 -8l^2 U(t)~cos(\theta)dy d\phi -(t^2+l^2)d\theta^2 
~~~~~~~~~~~~~\nonumber\\
~~~~~~~~~~~~~+(-4l^2U(t)~cos(\theta)^2-(t^2+l^2)~sin(\theta)^2)d\phi^2 
+ \frac{1}{U(t)}dt^2\nonumber
\end{eqnarray}
the coordinates being $(y,\theta,\phi,t)$. We obtained the Dirac equation as :
\begin{eqnarray}
\frac{i\hbar}{t^2+l^2}\left[-\gamma^0\left(\frac{t^2+l^2}{4\sqrt{U(t)}}
U^{\prime}(t) + \sqrt{U(t)}\left(1 +\frac{\partial}{\partial t}\right)\right)-
cotg(\theta)\sqrt{t^2+l^2}\gamma^2\right]\psi(t) \nonumber\\
- mc\psi(t)=0\nonumber
\end{eqnarray}
~~~~~~~~~\\

\noindent {\bf 3.} {\it G\" odel spacetime}, having the line element as :
\begin{eqnarray}
ds^2 = -a^2 dx^2 +\frac{1}{2}a^2 e^{2x}dy^2 +2a^2 e^{x}cdy dt -a^2 dz^2 +a^2 
c^2dt^2
\nonumber
\end{eqnarray}
in coordinates $(x,y,z,t)$. Here the Dirac equation is simply :
\begin{eqnarray}
i\hbar\frac{1}{a}\left [ - \gamma^1\left (\frac{1}{2}+\frac{\partial}{\partial
x}\right)+\sqrt{2}\gamma^2\left(e^{-x}\frac{\partial}{\partial y} -\frac{1}{c}
\frac{\partial}{\partial t}\right)-\gamma^3\frac{\partial}{\partial z}+
\frac{1}{c}\frac{\partial}{\partial t}\right]\psi(x,y,z,t) \nonumber\\
- mc\psi(x,y,z,t)=0
\nonumber
\end{eqnarray}

\noindent {\bf 4.} {\it McCrea static spacetime} \cite{6} with {\bf torsion} 
having the line element as
\begin{eqnarray}
ds^2 = -e^{2f(x)}dt^2 + dx^2 + e^{2g(x)}dy^2 + e^{2h(x)}dz^2\nonumber
\end{eqnarray}
in $(x,y,z,t)$ coordinates. If the coordinate lines of $y$ are closed with
$0\leq y < 2\pi$ and $-\infty < z < \infty$, $ 0 < x < \infty$, the spacetime
is cylindrically with $y$ as the angular, $x$ the cylindrical radial and $z$ 
the longitudinal coordinate. If $-\infty , x,y,z < \infty$ the symmetry is pseudo-planar. In \cite{6} 
McCrea considers the simplest solution of Einstein equations
with cosmological constant as
\begin{equation}\label{Mc}
f=h=h=qx/3
\end{equation}
and the cosmological constant turns to be $q^2/3$. We shall first consider
the general case specializing the results at the final step of the program 
to the above particular solution. Running our MAPLE+GRTensorII procedures, at
first we obtain the torsion tensor component as :
\begin{eqnarray}
{T}^y_{yx} = \frac{\partial v2(x)}{\partial x}e^{v2(x)}\hbox{~;~~}
{T}^z_{zx} = \frac{\partial v3(x)}{\partial x}e^{v3(x)}\hbox{~;~~}
{T}^t_{tx} = \frac{\partial v4(x)}{\partial x}e^{v4(x)}\nonumber
\end{eqnarray}
the rest of the components being zero. This time we have obtained the Dirac 
equation, depending also on the components of the torsion tensor as :
\begin{eqnarray}
i\hbar \frac{1}{2}\gamma^1\left ( e^{v2(x)}\frac{\partial v2(x)}{\partial x}+
e^{v3(x)}\frac{\partial v3(x)}{\partial x}+e^{v4(x)}\frac{\partial v4(x)}
{\partial x}-\frac{\partial f(x)}{\partial x}-\frac{\partial g(x)}{\partial x}
\right. \nonumber\\
\left.
-\frac{\partial h(x)}{\partial x} +2\frac{\partial}{\partial x}\right)\psi(x)-
mc\psi(x)=0\nonumber
\end{eqnarray}
Of course when we specialize to the particular solution proposed by McCrea in
\cite{6} we have to assign the form of metric functions as in (\ref{Mc}) and we can
then take the torsion functions as
\begin{eqnarray}
v2=v3=v4=v(x)\nonumber
\end{eqnarray}
and the Dirac equation becomes
\begin{eqnarray}
i\hbar\gamma^1\left(\frac{3}{2}\frac{\partial v(x)}{\partial x}e^{v(x)}-
\frac{1}{2}q + \frac{\partial}{\partial x}\right)\psi(x) -mc\psi(x)=0\nonumber
\end{eqnarray}

\noindent {\bf 5} {\it Schwarzschild metric} with {\bf torsion}. This example
is interesting in the view of recent investigations on the contribution of
torsion in gravity experiments using atomic interferometry \cite{10}. 
Here we used
the Schwarzschild metric having the line element written as
\begin{eqnarray}
ds^2 = e^{\lambda(r)}dr^2 +r^2d\theta^2 +r^2~sin(\theta)^2 d\phi^2 -
e^{\nu(r)}dt^2 \nonumber
\end{eqnarray}
Here we prefer to specialize the form of $\lambda(r)$ and $\nu(r)$ functions
as
\begin{eqnarray}
\nu(r) = 1-\frac{2m}{r}=\frac{1}{\lambda(r)}\nonumber
\end{eqnarray}
(we have $G=c=1$) later, after obtaining the form of the Dirac equation
in term of $\lambda(r)$ and $\nu(r)$ functions.

Using the same ``trick'' as in the previous example, we choose the components
of the torsion tensor as
\begin{eqnarray}
{T}^r_{r r} = \frac{1}{2}\frac{\partial f1(r)}{\partial r}e^{1/2 f1(r)}
\hbox{~;~~}
{T}^{\theta}_{\theta r}=1
\hbox{~;~~}
{T}^{\phi}_{\phi r}=sin(\theta)\nonumber
\end{eqnarray}
\begin{eqnarray}
{T}^t_{t r} = \frac{1}{2}\frac{\partial f2(r)}{\partial r}e^{1/2 f2(r)}
\hbox{~;~~}
{T}^{\phi}_{\phi \theta} = r~cos(\theta)
\nonumber
\end{eqnarray}
the rest of the components being zero. Running away our procedures we obtain
the Dirac equation containing terms with torsion tensor components as :
\begin{eqnarray}
i\hbar\left[ \frac{1}{4}e^{-\lambda(r)/2}\gamma^1
\left(\frac{\partial f2(r)}{\partial 
r}e^{f2(r)/2} - \frac{\partial \nu(r)}{\partial r} -\frac{4}{r} +2(1+sin(
\theta))+4\frac{\partial}{\partial r}\right) +\right .\nonumber\\
\left .\frac{1}{2}\gamma^2(cos(\theta)-
cotg(\theta))\right]\psi(r) - m\psi(r)=0\nonumber
\end{eqnarray}
  
\section*{Acknowledgments}

The author wish to thank  E. Seidel, D. Pollney, M. Alcubierre and many
others, for help, support and fruitful discussions. Special thanks go
to A. Popova, for carefully and critically reading of the article.

\end{document}